# Impact of strain and field ramp functional form on thermomagnetic instabilities in composite Nb$_3$Sn wires with multi-filaments inside the superconducting coil

Qing-Yu Wang[1], Jian-Bo Li[2], An He[2], Wei Liu[3], Cun Xue[4]*, You-He Zhou[5]*

[1]*School of Aeronautics, Northwestern Polytechnical University, Xi'an 710072, China*

[2]*College of Science, Chang'an University, Xi'an 710064, China*

[3]*Western Superconducting Technologies Co., Ltd. Xi'an 710014, China*

[4]*School of Mechanics, Civil Engineering and Architecture, Northwestern Polytechnical University, Xi'an 710072, China*

[5]*Key Laboratory of Mechanics on Disaster and Environment in Western China Attached to the Ministry of Education of China and Department of Mechanics and Engineering Sciences, Lanzhou University, Lanzhou 730000, China*

## Abstract

Superconducting Nb$_3$Sn magnets have composite structures consisting of Nb$_3$Sn multi-filaments, cooper, and epoxy resin. The complicated strain states due to different thermal expansion coefficients for Nb$_3$Sn and matrix materials and high Lorentz force will cause significant degradation of critical current density $J_c$ of Nb$_3$Sn wires, thereby having a substantial impact on the thermomagnetic instabilities in the superconducting magnets. So, it is highly needed to explore the effects of strain on thermomagnetic instabilities with numerical simulations. We theoretically analyze the effects of Cu/SC (superconductor) ratio, strain, and the field ramp functional form on flux jumps in composite Nb$_3$Sn wires inside the superconducting coil to obtain more accurate results. Considering the composite multi-filamentary structures, we find that a lower Cu/SC ratio leads to higher temperature peaks, and a larger proportion of Cu causes higher voltage peaks. Moreover, the strain significantly causes a higher frequency of flux jumps and higher voltage peaks. The temperatures recover to working temperature more difficultly, and SC wires quench earlier in the presence of strain. For the ramping magnetic field with a jagged ramp form, it is interesting that few flux jumps and temperature jumps occur at the decreasing branch, whereas frequent flux jumps can be observed promptly again when the applied current exceeds the pre-existing peak value. Our simulated results agree very well with experimental observations in Nb$_3$Sn coils. Additionally, unlike the pulsed flux jumps observed in linear ramp cases, giant and prolonged flux jumps are observed at the increasing branch under linear field ramp with additional sinusoidal field oscillations, when the applied current is sufficiently large. Reverse voltage signals are also observed during a decreasing branch, which can be revealed by the variations of current, magnetic field, and temperature distributions. The findings in this paper provide new insights into understanding and exploring the complex flux jumps in composite Nb$_3$Sn wires with multi-filaments.

* Corresponding authors: xuecun@nwpu.edu.cn; zhouyh@lzu.edu.cn.

# I. Introduction

Nb$_3$Sn with the A15 structure was discovered in 1954 and became an essential intermetallic compound due to its good superconductivity in extremely low-temperature regions[1-4]. With high critical current density $J_c$ [4, 5] in high fields (up to 3000 A/mm$^2$ at 12 T and 4.2 K), Nb$_3$Sn superconductors are widely used in particle accelerators, nuclear magnetic resonance (NMR), nuclear fusion, and magnetic resonance imaging (MRI) in the form of superconducting(SC) magnets[6-8]. For applications of Nb$_3$Sn in superconducting magnets, it is usually made of multi-filamentary wires embedded in a copper matrix [7, 9-11] to fulfill the desired engineering characteristics [12-15].

However, the composite Nb$_3$Sn wire is subjected to multiple strain states because of the thermal mismatch, manufacture and fabrication process, and high Lorentz force ($\mathbf{F_L} = \mathbf{J} \times \mathbf{B}$). Numerous investigations [16-24] have revealed that the strain sensitivity of Nb$_3$Sn is more than one order of magnitude higher than in other Nb-based superconductors such as NbTi and Nb, and the strain strongly reduces the critical current density of Nb$_3$Sn. To figure out the relationship between the strain and the critical current density $J_c$, Ekin designed the first experimental setup to test the dependence of critical current on strains in the 1980s[18]. Since then, plenty of experiments [25-31] has been done to investigate the effects of strains on the critical current density. Lots of theoretical and numerical research have been done to characterize the strain-dependent physical properties of composite Nb$_3$Sn wires based on constitutive models [4, 18, 21, 32-37], finite element models (FEM) [38-41], Ginzburg-Landau (GL) theory [42], and density functional theory (DFT) [43-45]. The severe degradation in the critical current density $J_c$ of Nb$_3$Sn caused the thermomagnetic instability or great destruction in the superconducting magnet [46].

Except for high critical current density $J_c$, the stable operation of superconducting magnets also demands other performance targets, such as combating instability, reducing field-ramp-rate-dependent ac loss, and reducing magnetic field induced by screening current[7, 11]. All the performance targets require a small filamentary size. However, the price for a small filamentary size is a relatively low critical current [7]. The filamentary size also has a considerable influence on thermomagnetic instabilities during ramping.

The maximal strain of our Nb$_3$Sn superconducting coil with an applied current of 700 A is about 0.9% which causes a severe decrease in critical current. Although the sensitivities of thermomagnetic instabilities to various electromagnetic and thermal parameters have been studied in experiments and numerical simulations [47-61], the effects of strain, filamentary size, and the field ramp functional form on the flux jump are still unclear. These parameters play a crucial role in the flux jump, so it is essential to reveal their influences on the superconducting magnets during the operation.

In this work, we use the representative unit cell (RUC) multi-physics numerical model [47] to numerically explore the sensitivity of flux jump to Cu/SC ratio, strain, and the field ramp functional form. We consider the composite multi-filamentary structures of $Nb_3Sn$ wires, the strains resulting from the thermal mismatch and high Lorentz force, and the applied field with a linear field ramp form, jagged field ramp form, or linear field ramp with additional sinusoidal field oscillations. We analyze the remarkable differences between the equivalent and multi-filamentary structures of superconducting wires. Our numerical results indicate that the strain and field ramp functional form significantly impact the thermomagnetic instabilities in $Nb_3Sn$ superconducting wires.

## II. Model

### 2.1 Framework of numerical model

At first, we briefly introduce our numerical model. As shown in Fig. 1, the transport currents strictly flow along the composite $Nb_3Sn$ wires without a short circuit and produce the surrounding magnetic fields. Based on the field profile calculated by Biot Savart's law, we consider the angular component of magnetic field $B_{//}$ as 0 on the condition that the transversal field $B_a$ is much larger than $B_{//}$. A representative quadrate unit cell which includes epoxy and one turn of the superconducting wires, is selected to analyze the thermomagnetic instabilities in composite $Nb_3Sn$ wires. The three-dimensional model is reduced to the two-dimensional axisymmetric model based on the symmetry of the quadrate unit cell. The diameter of the $Nb_3Sn$ superconducting wire $d_{sc}$ is 1.3 mm, and the cross-sectional area of the quadrate unit cell is 1.43×1.43 $mm^2$ (see Fig. 1). The strain distribution under a transport current of 700 A is obtained by the bidirectional homogenization method [62]. The maximum strain is 0.90 % at the inner radius, and the minimum strain is 0.64 % at the outer one, which is attributed to larger Lorentz force at the inner radius than at the outer one in the superconducting coil.

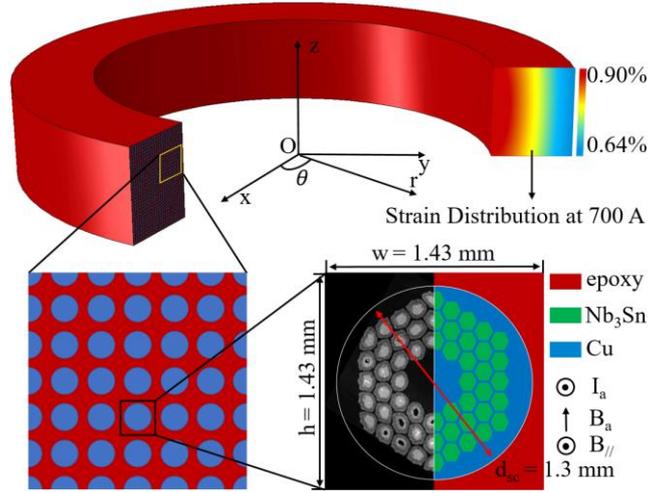

**Figure 1**. The schematic diagram and the representative unit cell of the superconducting coil.

**2.2 Electromagnetic field and heat conduction**

The *A-V* formulation [63, 64] of magnetoquasistatic Maxwell equations is employed to describe the electromagnetic properties of the SC wire. Here, we give a brief introduction to the *A-V* formulation. The electric field of an axially symmetric system at one point can be expressed as:

$$\mathbf{E} = -\frac{\partial \mathbf{A}^{\text{ext}}}{\partial t} - \frac{\partial \mathbf{A}^J}{\partial t} - \nabla V + \frac{\partial \Phi}{\partial t} \mathbf{e}_\theta \tag{1}$$

$$\nabla \times \mathbf{A}^{\text{ext}} = \mathbf{B}_a \tag{2}$$

$$V = v \frac{\theta}{2\pi} \tag{3}$$

$\mathbf{A}^{\text{ext}}$ is due to the external field. $\mathbf{A}^J$ is produced by the screening current in the superconducting wire. $V$ is characterized by the voltage per circle $v$ and the radian $\theta$. $\Phi$ is the flux. $\mathbf{e}_\theta$ is the unit vector in the circumferential direction. We simplify the solution of the three-dimensional Poisson equation $\nabla^2 \mathbf{A}^J = -\mu_0 \mathbf{J}$ to the axisymmetric cases [65] and assemble equations (1) - (3). The state equation can be obtained as follows:

$$\frac{\mu_0}{4\pi} \int_{S_i} \frac{\partial}{\partial t} J(r', z') \Gamma_{\text{as}}(r, z, r', z') dr' dz' = -E(r, z) - \frac{\partial A^{\text{ext}}(r, z)}{\partial t} - v + \frac{\partial \Phi}{\partial t} \tag{4}$$

We directly discretize the equation (4). The superconducting region is uniformly subdivided into N quadrate elements. The area of the *i*th subdivision is $S_i$. The state equation and constitutive equation for the *i*th element are defined as:

$$R(\mathbf{r}_i) I(\mathbf{r}_i) + \frac{dA^{\text{ext}}(\mathbf{r}_i)}{dt} + \sum_j M_{ij} \frac{d}{dt} I(\mathbf{r}_i) - \frac{d\Phi(\mathbf{r}_i)}{dt} = -v(\mathbf{r}_i) \tag{5}$$

$$A^{\text{ext}}(\mathbf{r}_i) = \frac{1}{S_i}\int_{S_i} A^{\text{ext}} dS \tag{6}$$

$$M_{ij} = \frac{\mu_0}{4\pi}\int_{S_i}\int_{S_j} \Gamma_{\text{as}}(\mathbf{r}_i,\mathbf{r}_j) dS dS' \tag{7}$$

$$R(\mathbf{r}_i) = \begin{cases} \dfrac{E_c}{J_{c0}} e^{\left(-\frac{U_0}{kT(\mathbf{r}_i)}\right)} e^{\left(\frac{|J(\mathbf{r}_i)|}{J_c(\mathbf{r}_i)}\frac{U_0}{kT(\mathbf{r}_i)}\right)} & T(\mathbf{r}_i)<T_c, B(\mathbf{r}_i)<B_c(\mathbf{r}_i), J(\mathbf{r}_i)<J_c(\mathbf{r}_i) \\ \dfrac{E_c}{J_{c0}} & \text{others} \end{cases} \tag{8}$$

$$\frac{U_0}{kT(\mathbf{r}_i)} = n\left(1-\frac{T(\mathbf{r}_i)^4}{T_c^4}\right)\left(1-\frac{|B(\mathbf{r}_i)|}{B_c(\mathbf{r}_i)}\frac{T_c^2}{T_c^2-T(\mathbf{r}_i)^2}\right) \tag{9}$$

$$J_c(\mathbf{r}_i) = J_{c0}\frac{T_c-T(\mathbf{r}_i)}{T_c-T_0}\exp\left(-\frac{|B(\mathbf{r}_i)|}{B_0}\right) f(\varepsilon(\mathbf{r}_i)) \tag{10}$$

$E_c$ is the critical electric field. $n$ is characterized by flux creep exponent $n_0$, temperature $T$, and critical temperature $T_c$. $B_c$ is the upper critical magnetic field when $T=0$ K. $J_c$ is the critical current density characterized by $T$, $B$, and $\varepsilon$, where $\varepsilon$ is the strain along the circumferential direction. In one superconducting wire, the total screen current induced by the external field is zero, which means that the filaments in the unit cell are fully coupled[66]. We assemble the state equations for each element.

$$\mathbf{M}\frac{d}{dt}\mathbf{I} = -\mathbf{R}\mathbf{I} - \frac{d}{dt}\mathbf{A}^{\text{ext}} + \frac{d}{dt}\boldsymbol{\Phi} - \mathbf{1}v \tag{11}$$

$$\mathbf{1}^{\text{T}}\mathbf{I} = I_a \tag{12}$$

We can obtain the field and current distributions in the unit cell by equations (11) and (12).

To obtain the temperature distribution of unit cell at any time, the governing equations of heat diffusion are given by:

$$c\frac{\partial T}{\partial t} = \nabla\cdot(\lambda\nabla T) - h(T-T_0) + \mathbf{E}\cdot\mathbf{J} \tag{13}$$

The specific heat $c$ and the thermal conductivity $\lambda$ are temperature-dependent. $h$ is nonzero only at the edges of the unit cell, which approximatively denotes the capabilities to take heat away. We assume that $h$, $c$, and $\lambda$ are all proportional to $T^3$ [67].

## 2.3 Mechanical equivalent treatment and Physical parameters

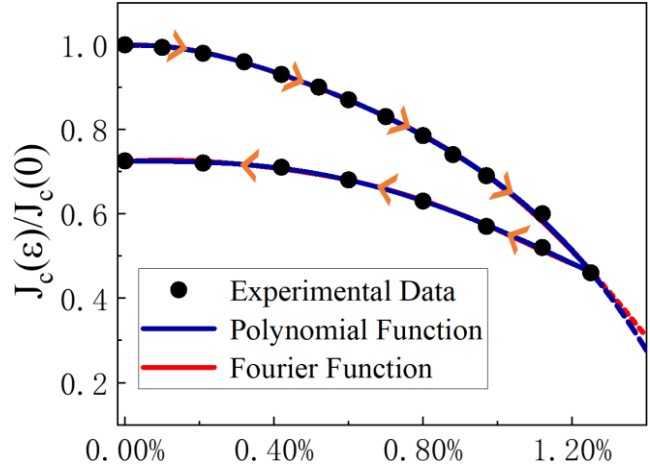

**Figure 2.** Strain dependence critical current density under loading and recovery (reprinted with permission from Ref. [68]), and two fitting curves.

The experimental data of critical current density versus strain for Oxford wires are obtained in Ref. [68], as shown in Fig. 2. The fourth-order Polynomial and second-order Fourier functions are used in the curve-fitting. We choose the fourth-order Polynomial function, and $f(\varepsilon)$ under loading and recovery can be expressed as:

$$f(\varepsilon) = -2.81 \times 10^7 \varepsilon^4 + 5.95 \times 10^5 \varepsilon^3 - 6783\varepsilon^2 + 3.88\varepsilon + 1.00 \tag{16}$$

$$f(\varepsilon) = 1.09 \times 10^7 \varepsilon^4 - 2.70 \times 10^5 \varepsilon^3 - 9.28\varepsilon^2 - 0.0054\varepsilon + 0.72 \tag{17}$$

Considering the representative unit cell contain all the components ($Nb_3Sn$, Cu, Epoxy), we take the superconducting coil as a geometric configuration which array the unit cell alone the *r* and *z* directions. So, the representative volume element (RVE) method is more effective than fractional volume method and self-consistent method in calculating the anisotropic equivalent mechanical parameters of the superconducting coil. At first, we use finite element method to establish a representative unit cell model, that each component material is isotropic. Then, the stress and strain distribution of RVE under different displacements can be obtained by applying unit axial and unit shear strain, respectively. The anisotropic equivalent mechanical parameters of the superconducting coil can be calculated by taking the mean stresses and strains under different strain states into the constitutive relation of RVE. The consistency of mechanical response under equivalent parameters and actual material parameters validates the reliability of equivalent parameters. We calculate the displacement of the equivalent superconducting coil under the Lorentz force. We select the displacement data of several unit cells and re-apply these data to the actual composite structure to obtain the detailed strain distribution at sub-regions [62]. We assume that the equivalent material has always been in the elastic stage, and there exists no fracture of filaments during ramping.

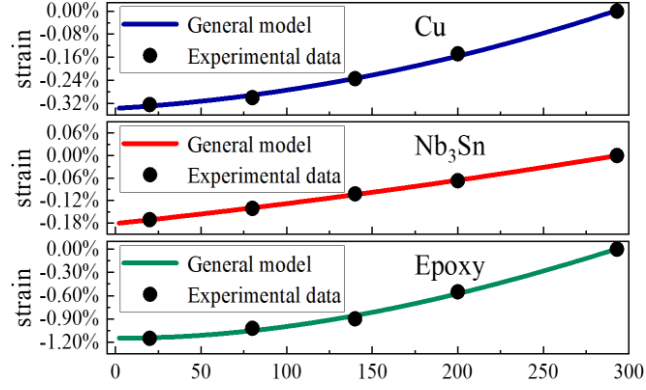

**Figure 3.** Experimental data of strain versus temperature for Cu, Nb₃Sn, and epoxy, and three fitting curves.

Fig. 3 shows the strains versus temperature for different materials [69]. Due to the mechanical equivalent treatment for the superconducting coil, the equivalent strain result from thermal expansion can be obtained:

$$\varepsilon_T(T) = \varepsilon_{SC}(T) \cdot V_{SC} + \varepsilon_{Cu}(T) \cdot V_{Cu} + \varepsilon_{epoxy}(T) \cdot V_{epoxy} \tag{18}$$

$V_{SC}$, $V_{Cu}$, and $V_{epoxy}$ represent the volume fractions of different materials, respectively. Note that the actual strain $\varepsilon$ in our simulation contains two parts.

$$\varepsilon = \varepsilon_T + \varepsilon_L \tag{19}$$

$\varepsilon_L$ results from the Lorentz force. The electromagnetic and thermal parameters for various kinds of materials are as follows. For Nb₃Sn, we adopt $I_{c0}$ = 10500 A, $B_{c2}$ = 28 T, $T_c$ = 18.2 K, $E_c$ = 0.0001 V/m, $B_0$ = 6.0 T, $\lambda_0$ = 0.1 W/m·K, $c_0$ = 1200 J/m³·K [48], $n_0$ = 20 and the maximum of $n(T) = n_0(\frac{T_c}{T})$ is limited as 100[70]. We set $\lambda_0$ = 300 W/m·K and $c_0$ = 660 J/m³·K for Cu [69], respectively. For epoxy, $\lambda_0$ = 0.1 W/m·K [71-73], $c_0$ = 400 J/m³·K and $h_0$ = 200 W/m³·K.

## III. Results and discussions

### 3.1 Effect of strain on flux jump in multi-filamentary Nb₃Sn wire

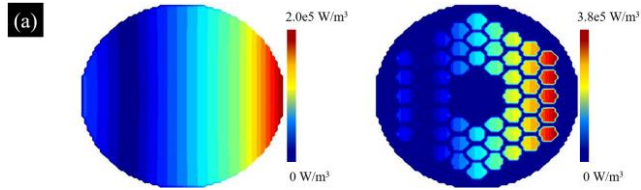

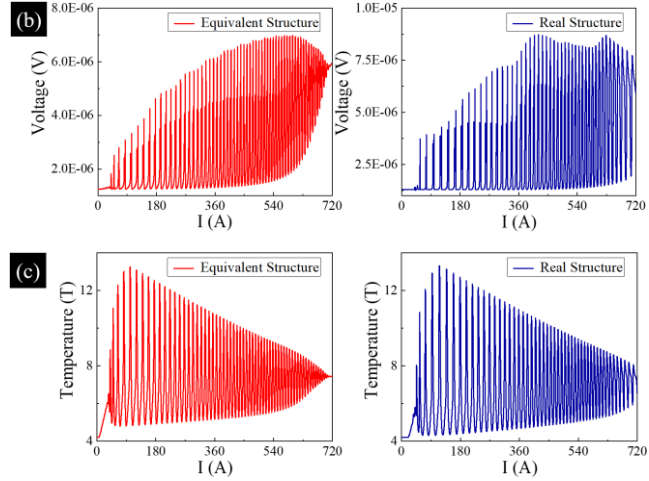

**Figure 4.** The contour plots of heating power (a), flux jumps (b), and temperature jumps (c) for the equivalent structure and the multi-filamentary structure with a ramp rate of 1 A/s, field ramp rate of 0.015 T/s, and $h_0$ = 100 W/m³·K.

At first, we focus on the difference between the equivalent structure and the actual structure. Note that Nb₃Sn and epoxy are the only two materials for the equivalent structure. Both the two structures have the same $I_{c0}$. The thermal conductivity and the specific heat for the equivalent wire are given by

$$\lambda(T) = \lambda_{Nb_3Sn}(T) \cdot S_{Nb_3Sn} + \lambda_{Cu}(T) \cdot S_{Cu} \qquad (20)$$

$$c(T) = c_{Nb_3Sn}(T) \cdot S_{Nb_3Sn} + c_{Cu}(T) \cdot S_{Cu} \qquad (21)$$

$S_{Nb3Sn}$ and $S_{Cu}$ represent the percentages of Nb₃Sn and Cu in the SC wire, respectively. There are apparent differences in the power profile, the voltage, and the temperature during operation, as shown in Fig. 4. Noticeable heating can be observed all over the equivalent SC wire during ramping, while it can only be observed in the filaments for the multi-filamentary structure (see Fig. 4(a)). The immense power of the equivalent structure leads to a rapid temperature rise (see Fig. 4(c)) compared with the actual structure. The rapid temperature rise narrows the range of flux jumps, which is attributed to the that a small flux creep exponent at the high temperature promotes flux creep. As shown in Fig. 4(b), the regularity of voltage jumps of the actual structure is different from the equivalent structure. The frequent flux jump in the equivalent structure (9 peaks before 180 A, while 7 peaks before 180 A for the actual structure) results in minor magnetic pressure build-up and small amplitudes of the voltages during ramping. There is no doubt that it is vital to investigate the flux jump in the actual structure based on the above results.

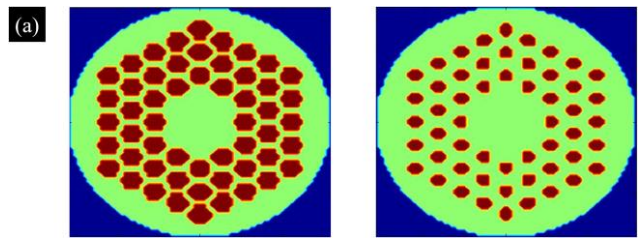

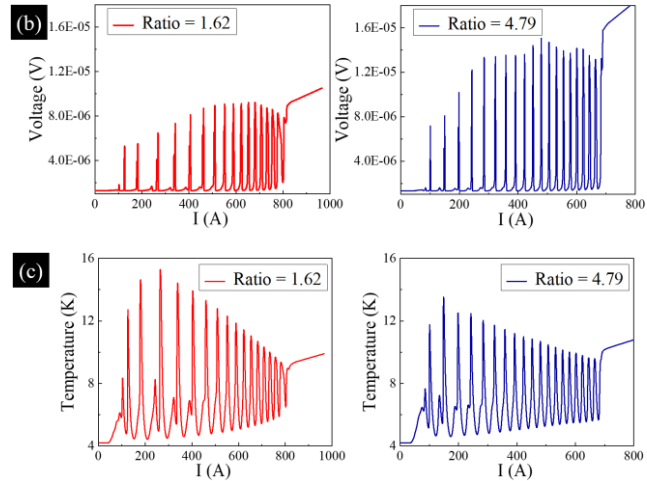

**Figure 5.** The sketches for structures with different Cu/SC ratios (1.62 and 4.79) (a), flux jumps (b), and temperature jumps (c) with a ramp rate of 1 A/s and field ramp rate of 0.015 T/s.

Although a high Cu/SC ratio leads to small filamentary size and reduction of ac losses, the price is a relatively low critical current. It causes problems for the superconductor application, which requires carrying a high transport current at high magnetic fields. However, a large filamentary size corresponding to a low Cu/SC ratio is harmful to $Nb_3Sn$ stability, ac loss reduction, and reduction of the magnetic field induced by shield current [7]. To figure out the characteristics of the thermomagnetic instabilities in the SC wire with different Cu/SC ratios, Fig. 5 shows the flux jumps in the wire close to the edge by varying the magnitude of filamentary size. The ramp rate is 1 A/s, and the field ramp rate is 0.015 T/s. We investigate two kinds of wire (Cu/SC ratio is 1.62 and 4.79). The sketches of the SC wire are illustrated in Fig. 5(a). The capability to carry high-loss current decreases with increasing Cu/SC ratio (see Fig. 5(b)) due to the decrease of the $Nb_3Sn$ fraction as the Cu fraction increases. The amplitudes of voltage peaks are higher in the Cu/SC ratio of 4.79. The frequency of flux jumps decreases with SC content, and the amplitude of temperature peaks increases with SC content (see Fig. 5(c)). These are attributed to the that high ac loss at large SC fraction causes rapid temperature rise, then restrains the flux jump.

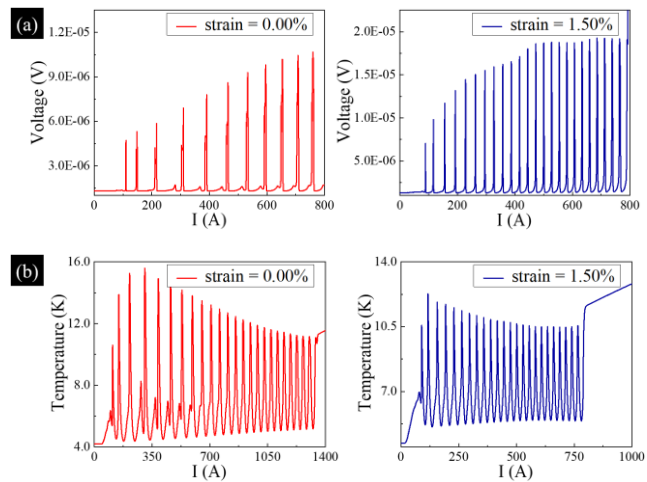

**Figure 6.** The flux jumps (a) and temperature jumps (b) in the absence or presence of a constant strain of 1.5 % with a ramp rate of 1 A/s and field ramp rate of 0.0064 T/s.

In the actual operation of superconducting magnets, considerable strains due to the fabrication process cause significant degradation in the superconductivity. The critical current density $J_c$ reduces nonlinearly and reversibly within certain limits. Therefore, it is vital to understand the effect of the reduction of $J_c$ on flux jump. We chose the SC wire located in the center of the superconducting coil with $h_0$ of 60 W/m³·K. The ramp rate is 1 A/s, and the field ramp rate is 0.0064 T/s. We assume a fixed strain result from the fabrication process and thermal mismatch on the SC wire. Fig. 6 shows the considerable difference between cases where the effect of strain on $J_c$ is considered or not. Both larger amplitudes of voltages peaks and higher frequency of flux jumps are observed in the SC wire subjected to strain (see Fig. 6(a)), which is attributed to the reduction of $J_c$ requires less steep magnetic field attenuation so that more flux vortices penetrate into the SC wire. As the frequent flux jumps, the temperature is more challenging to return to the working temperature. As illustrated in Fig. 6(b), the quench current decreases from 1400 A to 800 A because of the strain. The strain restrains the temperature peaks because of the reduction of $J_c$, which causes lower heating power. It is not appropriate for any analysis of flux jumps in magnets exposed to high currents to ignore the influence of strain.

### 3.2 Effects of field ramp functional form on flux jump in the presence of strain

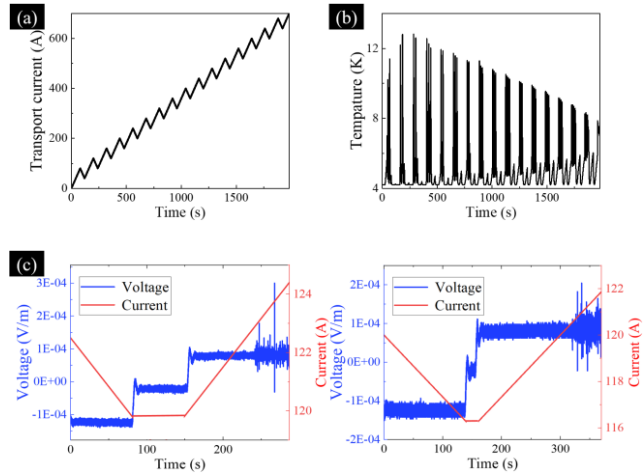

**Figure 7.** The ramp functional form (a) where the applied current increases by 80 A and then decreases by 40 A with ramp rate |1 A/s| and corresponding temperature jumps (b) in the presence of current-dependent strain. The partial experimental data with a jagged field ramp form (c).

Except for the effects of the manufacture and fabrication process, the thermal contraction differences between materials on superconducting properties, the Lorentz forces during operation also are considered. We investigate the thermomagnetic instabilities in SC wires with a jagged field ramp form. Fig. 7(a) shows that the transport current along with the SC wire increases by 80 A and then decreases by 40 A with a ramp rate of 1 A/s. When the reduction of transport current in decreasing branches is as small as a specific value, Fig. 7(b) indicates that concentrated flux jumps occur as the current exceeds the previous current peaks, and almost few flux jumps and corresponding temperature jumps occur as the applied current decrease. Our numerical results are consistent with the experiment results (see Fig. 7(c)) observed in $Nb_3Sn$ coils. Compared to the case with the linear field ramp form in Fig. 6(b), the temperature can recover to the ambient temperature after flux jumps, promoting the superconducting magnet's stable operation and high capacity to carry current. It also inspires us to optimize the operation of superconducting magnets from the field ramp functional form.

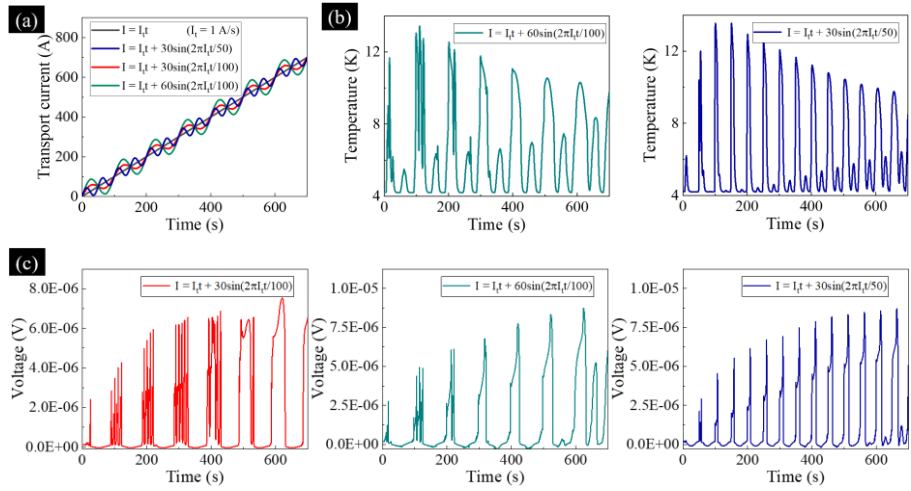

**Figure 8.** The different ramp functional forms (a), corresponding simulated temperature jumps (b), and flux jumps (c) in the presence of current-dependent strain.

To figure out the influence of field ramp functional form on thermomagnetic instabilities in the presence of current-dependent strain, we consider linear field ramp with additional sinusoidal field oscillations and change both the amplitude and the frequency of the sine wave. The SC wire is close to the edge, and the ramp functional forms are shown in Fig. 8(a), where the transport current is increased from 0 A to 700 A. Remarkable differences are observed between Fig. 8(b) and Fig. 6(b). Giant and prolonged flux jumps are observed at the increasing branch when the applied current is sufficiently large. In contrast with the linear field ramp cases, the temperature in the SC wire can recover to working temperature more easily. Unlike the case in Fig. 6(b), dense flux jumps exist at the branches with a high field ramp rate corresponding to the rapid accumulation of magnetic pressure (see Fig. 8(c)). As illustrated in Fig. 8(c), the statistics of voltage peaks under linear field ramp with additional sinusoidal field oscillations are significantly less than those under a linear field ramp form. Additionally, we investigate the characteristics of thermomagnetic instabilities in the SC wire during decreasing branches. We consider the irreversible degradation of critical current density due to the strain. Fig. 9(a) illustrate the decreasing branches with different ramp rate (-1.0 A/s, -0.5 A/s and -0.25A/s). The $B_a/I_a$ and $h_0$ are 0.0064 T/A and 60 W/m$^3$·K, respectively. As shown in Fig. 9(b), the statistics of flux jumps during decreasing branch increase with decreasing field ramp rate (16 and 20). The temperature is easier to recover to the ambient temperature after flux jumps, so a slow field ramp rate promotes the protection of the superconducting magnet. The voltage peaks during increasing branches are generally larger than those during decreasing branches (see Fig. 9(c)). Both flux jumps and reverse flux jumps are observed when the applied current is reduced from 1000 A to 0 A. At the decreasing branch, the frozen flux in the inner regions [74] and the reverse temperature distribution in contrast with the cases in the increasing branch cause a lower $J_c$ region in the interior of the Nb$_3$Sn wire (see Fig. 9(d)). The reverse current profile and high resistivity region due to the lower $J_c$ lead to the reverse voltage signals, as shown in Fig. 9(c).

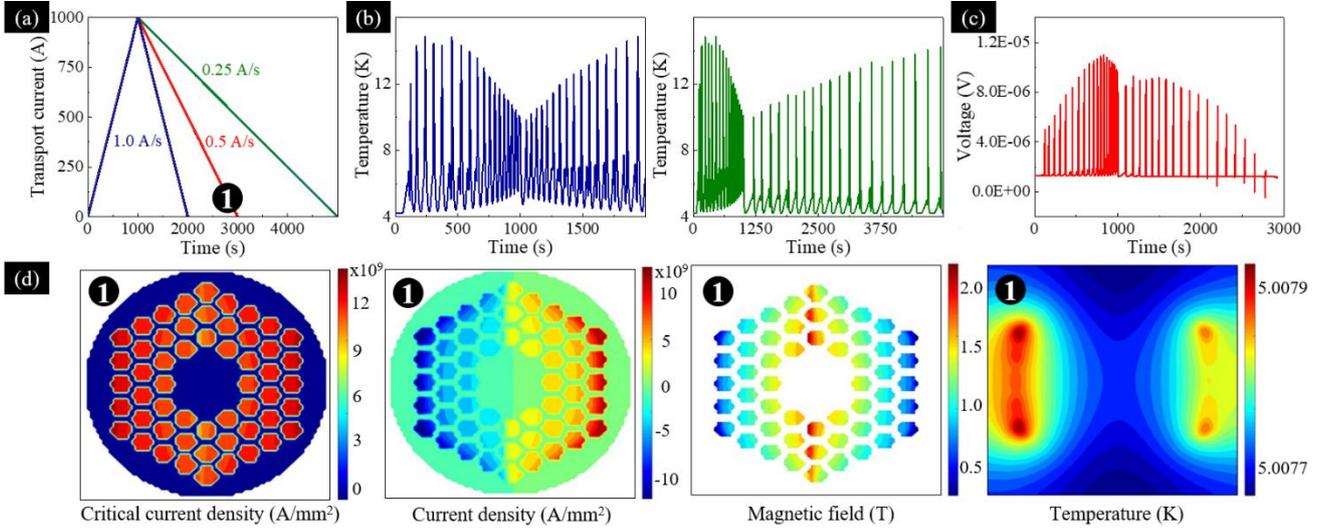

**Figure 9.** The different unloading ramp rates (a), temperature jumps (b), voltage jumps (c), and contour plots of the critical current density, current density, magnetic field, and temperature (d).

## IV. Conclusion

In this paper, an appropriate theoretical method mentioned in Ref. [47] is adopted to analyze the effects of Cu/SC ratio, strain, and the field ramp functional form on thermomagnetic instabilities of the superconducting coil. The approximate strain profile across the cross-section of the superconducting coil is obtained based on the equivalent mechanical model by simplifying the materials for the coil to an equivalent anisotropic material. We illustrate that the Cu/SC ratio, strain, and field ramp functional form play significant roles in thermomagnetic instabilities. A smaller Cu/SC ratio leads to higher temperature peaks, and a larger proportion of Cu causes high voltage peaks. Additionally, the strain causes a higher frequency of flux jumps and higher voltage peaks. The temperature in the SC wire is more challenging to return to the working temperature, and the SC wire quench earlier due to the existence of strain. For the ramping magnetic field with a jagged ramp form, few flux jumps and temperature jumps occur at decreasing branches, whereas frequent flux jumps can be observed promptly again when the applied current exceeds the pre-existing peak value. Our numerical results agree well with the experiments. Moreover, remarkable differences between the linear field ramp forms and the linear field ramp with additional sinusoidal field oscillations are observed. Giant and prolonged flux jumps are observed at the increasing branch when the applied current is sufficiently large. The statistics of voltage peaks under linear field ramp with additional sinusoidal field oscillations are significantly less than those under linear ones. At the decreasing branch, the frozen flux in the inner regions and the reverse temperature distribution in contrast with the cases in the increasing branch cause a lower $J_c$ region in the interior of the $Nb_3Sn$ wire. The reverse current profile and high resistivity region because of the lower $J_c$ lead to the reverse voltage signals. This numerical simulation qualitatively indicates that the thermomagnetic instabilities of the superconducting coil are strongly sensitive to Cu/SC ratio, strain, and the field ramp functional form. Future work will focus on a precise analysis of the superconducting coil.

## Acknowledgment:

Q.-Y. W. and C. X. acknowledge support from the National Natural Science Foundation of China (Grants Nos. 11972298 and 12011530143). W. L. acknowledges support from the Shaanxi Province Key Research and Development Program of China (Grant No. 0105679005) and the Natural Science Basic Research Plan in Shaanxi Province of China (Grant No. 007234930030).

## References:

[1]   T Takao, $Nb_3Al$ conductors for high-field applications, *Supercond. Sci. Technol.* 13 (2000) R101.
[2]   D Ciazynski, Review of $Nb_3Sn$ conductors for ITER, *Fusion Eng. Des.* 82 (2007) 488-497.
[3]   Z Cao, K Hu, M Pan, Z Huang, W Hu, H Wu*, et al.*, Phase transformation and mechanical stability of niobium aluminide ($Nb_3Al$) induced by high pressures, *J. Alloys Compd.* 869 (2021) 159278.
[4]   A Godeke, A review of the properties of $Nb_3Sn$ and their variation with A15 composition, morphology and strain state, *Supercond. Sci. Technol.* 19 (2006) R68-R80.
[5]   R Flükiger, D Uglietti, C Senatore, and F Buta, Microstructure, composition and critical current density of superconducting $Nb_3Sn$ wires, *Cryogenics* 48 (2008) 293-307.
[6]   S Posen and D L Hall, $Nb_3Sn$ superconducting radiofrequency cavities: fabrication, results, properties, and prospects, *Supercond. Sci. Technol.* 30 (2017) 033004.
[7]   X Xu, A review and prospects for $Nb_3Sn$ superconductor development, *Supercond. Sci. Technol.* 30 (2017) 093001.

[8] L Sun, W Lu, E M Mei, G Sabbi, D Xie, W Wu, *et al.*, Superconducting magnets for high performance ECR ion sources, *IEEE Trans. Appl. Supercond.* 28 (2018) 4101606.

[9] V V Kashikhin and A V Zlobin, Magnetic Instabilities in $Nb_3Sn$ Strands and Cables, *IEEE Trans. Appl. Supercond.* 15 (2005) 1621-1624.

[10] L D Cooley, S P Chang, and K A Ghosh, Magnetization, RRR and Stability of $Nb_3Sn$ Strands With High Sub-Element Number, *IEEE Trans. Appl. Supercond.* 17 (2007) 2706-2709.

[11] X Wang, G Ambrosio, G Chlachidze, E W Collings, D R Dietderich, J DiMarco, *et al.*, Validation of Finite-Element Models of Persistent-Current Effects in $Nb_3Sn$ Accelerator Magnets, *IEEE Trans. Appl. Supercond.* 25 (2015) 1-6.

[12] K Inoue, Y Iijima, and T Takeuchi, Superconducting properties of $Nb_3Al$ multifilamentary wire, *Appl. Phys. Lett.* 52 (1988) 1724-1725.

[13] T Takeuchi, K Tsuchiya, M Saeda, N Banno, A Kikuchi, and Y Iijima, Electron backscatter diffraction analysis of $Nb_3Al$ multifilamentary strands prepared by rapid heating, quenching and transformation annealing, *Supercond. Sci. Technol.* 23 (2010) 125001.

[14] C Scheuerlein, G Arnau, P Alknes, N Jimenez, B Bordini, A Ballarino, *et al.*, Texture in state-of-the-art $Nb_3Sn$ multifilamentary superconducting wires, *Supercond. Sci. Technol.* 27 (2014) 025013.

[15] A Godeke, M C Jewell, C M Fischer, A A Squitieri, P J Lee, and D C Larbalestier, The upper critical field of filamentary $Nb_3Sn$ conductors, *J. Appl. Phys.* 97 (2005) 093909.

[16] X-F Lu, S Pragnell, and D P Hampshire, Small reversible axial-strain window for the critical current of a high performance $Nb_3Sn$ superconducting strand, *Appl. Phys. Lett.* 91 (2007) 132512.

[17] J W Ekin, Effect of stress on the critical current of $Nb_3Sn$ multifilamentary composite wire, *Appl. Phys. Lett.* 29 (1976) 216-219.

[18] J W Ekin, Strain scaling law for flux pinning in practical superconductors. Part 1: Basic relationship and application to Nb3Sn conductors, *Cryogenics* 20 (1980) 611-624.

[19] J W Ekin, N Cheggour, L Goodrich, J Splett, B Bordini, and D Richter, Unified Scaling Law for flux pinning in practical superconductors: II. Parameter testing, scaling constants, and the Extrapolative Scaling Expression, *Supercond. Sci. Technol.* 29 (2016) 123002.

[20] X-F Lu, D M J Taylor, and D P Hampshire, Critical current scaling laws for advanced $Nb_3Sn$ superconducting strands for fusion applications with six free parameters, *Supercond. Sci. Technol.* 21 (2008) 105016.

[21] D F Valentinis, C Berthod, B Bordini, and L Rossi, A theory of the strain-dependent critical field in $Nb_3Sn$, based on anharmonic phonon generation, *Supercond. Sci. Technol.* 27 (2014) 025008.

[22] G De Marzi, V Corato, L Muzzi, A d Corte, G Mondonico, B Seeber, *et al.*, Reversible stress-induced anomalies in the strain function of $Nb_3Sn$ wires, *Supercond. Sci. Technol.* 25 (2012) 025015.

[23] G Rupp, Enhancement of the critical current of multifilamentary $Nb_3Sn$ conductors by tensile stress, *J. Appl. Phys.* 48 (1977) 3858-3863.

[24] J W Ekin, Strain scaling law for flux pinning in NbTi, $Nb_3Sn$, Nb-HfCu-Sn-Ga, $V_3Ga$ and $Nb_3Ge$., *IEEE Trans. Magn.* 17 (1981) 658-661.

[25] A Nijhuis, N C v d Eijnden, Y Ilyin, E G v Putten, G J T Veening, W A J Wessel, *et al.*, Impact of spatial periodic bending and load cycling on the critical current of a $Nb_3Sn$ strand, *Supercond. Sci. Technol.* 18 (2005) S273-S283.

[26] A Nijhuis, Y Ilyin, and W Abbas, Axial and transverse stress–strain characterization of the EU dipole high current density $Nb_3Sn$ strand, *Supercond. Sci. Technol.* 21 (2008) 065001.

[27] M Takayasu, L Chiesa, D L Harris, A Allegritti, and J V Minervini, Pure bending strains of $Nb_3Sn$ wires, *Supercond. Sci. Technol.* 24 (2011) 045012.

[28] J W Ekin, Effect of transverse compressive stress on the critical current and upper critical field of Nb$_3$Sn, *J. Appl. Phys.* 62 (1987) 4829-4834.

[29] L Chiesa, M Takayasu, J V Minervini, C Gung, P C Michael, V Fishman, *et al.*, Experimental Studies of Transverse Stress Effects on the Critical Current of a Sub-Sized Nb$_3$Sn Superconducting Cable, *IEEE Trans. Appl. Supercond.* 17 (2007) 1386-1389.

[30] B Seeber, A Ferreira, V Abächerli, and R Flükiger, Critical current of a Nb$_3$Sn bronze route conductor under uniaxial tensile and transverse compressive stress, *Supercond. Sci. Technol.* 20 (2007) S184-S188.

[31] A Nijhuis, R P Pompe van Meerdervoort, H J G Krooshoop, W A J Wessel, C Zhou, G Rolando, *et al.*, The effect of axial and transverse loading on the transport properties of ITER Nb$_3$Sn strands, *Supercond. Sci. Technol.* 26 (2013) 084004.

[32] W D Markiewicz, Elastic stiffness model for the critical temperature Tc of Nb$_3$Sn including strain dependence, *Cryogenics* 44 (2004) 767-782.

[33] W D Markiewicz, Comparison of strain scaling functions for the strain dependence of composite Nb$_3$Sn superconductors, *Supercond. Sci. Technol.* 21 (2008) 054004.

[34] S Oh and K Kim, A scaling law for the critical current of Nb$_3$Sn stands based on strong-coupling theory of superconductivity, *J. Appl. Phys.* 99 (2006) 255.

[35] D M J Taylor and D P Hampshire, The scaling law for the strain dependence of the critical current density in Nb$_3$Sn superconducting wires, *Supercond. Sci. Technol.* 18 (2005) S241-S252.

[36] J W Ekin, N Cheggour, L Goodrich, and J Splett, Unified Scaling Law for flux pinning in practical superconductors: III. Minimum datasets, core parameters, and application of the Extrapolative Scaling Expression, *Supercond. Sci. Technol.* 30 (2017) 033005.

[37] D-H Yue, X-Y Zhang, J Zhou, and Y-H Zhou, Current transport of the [001]-tilt low-angle grain boundary in high temperature superconductors, *Appl. Phys. Lett.* 103 (2013) 232602.

[38] S Murase, H Okamoto, T Wakasa, T Tsukii, and S Shimamoto, Three-directional analysis of thermally-induced strains for Nb$_3$Sn and oxide composite superconductors, *IEEE Trans. Appl. Supercond.* 13 (2003) 3386-3389.

[39] W Tiening, L Chiesa, and M Takayasu, An FE Model to Study the Strain State of the Filaments of a Nb$_3$Sn Internal-Tin Strand Under Transverse Load, *IEEE Trans. Appl. Supercond.* 23 (2013) 8400205-8400205.

[40] H Bajas, D Durville, and A Devred, Finite element modelling of cable-in-conduit conductors, *Supercond. Sci. Technol.* 25 (2012) 054019.

[41] X Wang, Y-X Li, and Y-W Gao, Mechanical behaviors of multi-filament twist superconducting strand under tensile and cyclic loading, *Cryogenics* 73 (2016) 14-24.

[42] F Xue and Y-H Zhou, "Effect of strain on critical current density in grain boundaries of superconductors," 2013.

[43] R Loria, G De Marzi, S Anzellini, L Muzzi, N Pompeo, F Gala, *et al.*, The Effect of Hydrostatic Pressure on the Superconducting and Structural Properties of Nb$_3$Sn: Ab-initio Modeling and SR-XRD Investigation, *IEEE Trans. Appl. Supercond.* 27 (2017) 1-5.

[44] G De Marzi, L Morici, L Muzzi, A della Corte, and M Buongiorno Nardelli, Strain sensitivity and superconducting properties of Nb$_3$Sn from first principles calculations, *J. Phys.: Condens. Matter* 25 (2013) 135702.

[45] R Zhang, P-F Gao, and X-Z Wang, Strain dependence of critical superconducting properties of Nb$_3$Sn with different intrinsic strains based on a semi-phenomenological approach, *Cryogenics* 86 (2017) 30-37.

[46] M. Bajko, et. al. "Report of the Task Force on the Incident of 19th September 2008 at the LHC", *LHC-PROJECT-Report1168, CERN,* 2009

[47] Q-Y Wang, C Xue, Y-Q Chen, X-J Ou, W Wu, W Liu, *et al.*, Thermomagnetic instabilities of Nb$_3$Sn wires inside the superconducting solenoid, *Physica C* 593 (2021) 1354002.

[48] X Xu, P Li, A V Zlobin, and X Peng, Improvement of stability of Nb$_3$Sn superconductors by introducing high specific heat substances, *Supercond. Sci. Technol.* 31 (2018) 03LT02.

[49] K Shiiki and M Kudo, Anomalous hysteresis loss in superconducting wire due to flux jump, *J. Appl. Phys.* 45 (1974) 4071-4075.

[50] A V Pan, S Zhou, H Liu, and S Dou, Properties of superconducting MgB$_2$ wires: in situ versus ex situ reaction technique, *Supercond. Sci. Technol.* 16 (2003) 639.

[51] T Takeuchi, K Tsuchiya, K Nakagawa, S Nimori, N Banno, Y Iijima, *et al.*, A new RHQT Nb$_3$Al superconducting wire with a Ta/Cu/Ta three-layer filament-barrier structure, *Supercond. Sci. Technol.* 25 (2012) 065016.

[52] A V Zlobin, V V Kashikhin, and E Barzi, Effect of Flux Jumps in Superconductor on Nb$_3$Sn Accelerator Magnet Performance, *IEEE Trans. Appl. Supercond.* 16 (2006) 1308-1311.

[53] J Coello de Portugal, R Tomás, L Fiscarelli, D Gamba, and M Martino, Impact of flux jumps in future colliders, *Phys. Rev. Accel. Beams* 23 (2020) 011001.

[54] B Bordini and L Rossi, Self Field Instability in High-Jc Nb$_3$Sn Strands With High Copper Residual Resistivity Ratio, *IEEE Trans. Apl. Supercond.* 19 (2009) 2470-2476.

[55] M Martino, P Arpaia, and S Ierardi, Impact of Flux Jumps on High-Precision Powering of Nb$_3$Sn Superconducting Magnets, *J. Phys.: Conf. Ser.* 1350 (2019) 012180.

[56] E Barzi, F Berritta, D Turrioni, and A V Zlobin, Heat Diffusion in High-C$_p$ Nb$_3$Sn Composite Superconducting Wires, *Instruments* 4 (2020) 28.

[57] V S Vysotsky, M Takayasu, S Jeong, P C Michael, and V V Vysotskaia, Voltage spikes in superconducting CableIn-Conduit Conductor under ramped magnetic fields. Part 2: Analysis of loop inductances and current variations associated with the spikes, *Cryogenics* 38 (1998) 387-395.

[58] L Jiang, C Xue, L Burger, B Vanderheyden, A V Silhanek, and Y-H Zhou, Selective triggering of magnetic flux avalanches by an edge indentation, *Phys. Rev. B* 101 (2020) 224505.

[59] S Jeong, J H Schultz, M Takayasu, V Vysotsky, P C Michael, W Warnes, *et al.*, Voltage spike observation in superconducting cable-in-conduit conductor under ramped magnetic fields: 1. Experiments, *Cryogenics* 37 (1997) 299-304.

[60] Y-H Zhou and X-B Yang, Numerical simulations of thermomagnetic instability in high-Tcsuperconductors: Dependence on sweep rate and ambient temperature, *Phys. Rev. B* 74 (2006) 054507.

[61] C-G Huang, B Xu, and Y-H Zhou, Dynamic simulations of actual superconducting maglev systems considering thermal and rotational effects, *Supercond. Sci. Technol.* 32 (2019) 045002.

[62] A He, J-B Li, and C Xue, Bidirectional homogenization method for accurate analysis of mechanical behaviors of Nb$_3$Sn superconducting coils, *Chinese Journal of Theoretical and Applied Mechanics* 54 (2022) 1274-1290.

[63] A Morandi, 2D electromagnetic modelling of superconductors, *Supercond. Sci. Technol.* 25 (2012) 104003.

[64] A Morandi and M Fabbri, A unified approach to the power law and the critical state modeling of superconductors in 2D, *Supercond. Sci. Technol.* 28 (2015) 024004.

[65] J. D. Jackson 1998 *Classical Electrodynamics 3rd edn* (New York: Wiley)

[66] V Lahtinen and A Stenvall, Toward Two-Dimensional Simulations of Hysteresis Losses in Partially Coupled Superconducting Wires, *IEEE Trans. Appl. Supercond.* 24 (2014) 1-5.

[67] J I Vestgården, D V Shantsev, Y M Galperin, and T H Johansen, Dynamics and morphology of dendritic flux avalanches in superconducting films, *Phys. Rev. B* 84 (2011) 054537.

[68] N Allen, P Mallon, J King, L Chiesa, and M Takayasu, Wide range pure bending strains of Nb$_3$Sn wires, *Supercond. Sci. Technol.* 27 (2014) 065014.

[69]    Y. Iwasa 2009 *Case Studies in Superconducting Magnets - Design and Operational Issues*, Plenum Press, New York

[70]    J I Vestgården, F Colauto, A M H de Andrade, A A M Oliveira, W A Ortiz, and T H Johansen, Cascade dynamics of thermomagnetic avalanches in superconducting films with holes, *Phys. Rev. B* 92 (2015) 144510.

[71]    F J Walker and A C Anderson, Thermal conductivity and specific heat of a glass–epoxy composite at temperatures below 4 K, *Rev. Sci. Instrum.* 52 (1981) 471-472.

[72]    R G Ross, Estimation of thermal conduction loads for structural supports of cryogenic spacecraft assemblies, *Cryogenics* 44 (2004) 421-424.

[73]    A L Woodcraft, V Martelli, and G Ventura, Thermal conductivity of ME771 glass–epoxy laminate from millikelvin temperatures to 4K, *Cryogenics* 50 (2010) 52-54.

[74]    C Xue, A He, H Yong, and Y Zhou, Field-dependent critical state of high-Tc superconducting strip simultaneously exposed to transport current and perpendicular magnetic field, *AIP Advances* 3 (2013) 122110.